\documentclass{IAU-TrA}
\usepackage{graphicx}

\title[INFRARED ASTRONOMY] 
{}

\author[DIVISION~XII / COMMISSION~50 ] 
{}

\pubyear{2012}
\volume{Volume XXVIIIA}
\pagerange{001--008}
\jname{Transactions IAU, Volume XXVIIIA}
\editors{Ian Corbett, ed.}
\begin{document}

\maketitle

{\bf

\large
\noindent
DIVISION XII / COMMISSION 50                             \\ 
PROTECTION OF EXISTING AND POTENTIAL OBSERVATORY SITES   \\

\normalsize

\begin{tabbing}
\hspace*{45mm}  \=                                                   \kill
CHAIR           \> Wim van Driel                                     \\
VICE-CHAIR      \> Richard Green                                     \\
PAST CHAIR      \> Richard J. Wainscoat                              \\
BOARD           \> Elizabeth Alvarez del Castillo                    \\   
                \> Carlo Blanco                                      \\  
                \> David L. Crawford                                 \\
                \> Margarita Metaxa                                  \\
                \> Masatoshi Ohishi                                  \\
                \> Woodruff T. Sullivan III                          \\
                \> Anastasios Tzioumis                               \\
                \end{tabbing}
\vspace{3mm}

\noindent
COMMISSION 50 WORKING GROUP \\
Div. XII / Commission 50 WG \,\,\, Controlling Light Pollution           \\


\vspace{3mm}

\noindent
TRIENNIAL REPORT 2009-2012
}

\firstsection 

\section{Introduction}
The activities of the Commission have continued to focus on controlling unwanted
light and radio emissions at observatory sites, monitoring of conditions at observatory
sites, and education and outreach. Commission members have been active in securing
new legislation in several locations to further the protection of observatory sites
as well as in the international regulation of the use of the radio spectrum and the
protection of radio astronomical observations.
\smallskip

To kick off the IAU/UNESCO International Year of Astronomy, 2009, IAU Symposium 260 on the 
R\^ole of Astronomy in Society and Culture was held in Paris. Commission 50 members gave
presentations on the importance of saving the magnificent night sky from light pollution and on the need to protect radio astronomy from unwanted interference.
\smallskip

Also in 2009, at its XXVII General Assembly in Rio de Janeiro, the IAU passed Resolution B5 
in Defence of the Night Sky and the Right to Starlight. The Resolution encourages IAU 
members to assist in raising public awareness about the contents and objectives of the International Conference in Defence of the Quality of the Night Sky and the Right to Observe Stars [http://www.starlight2007.net/], in particular the importance of preserving access to an unpolluted night sky for all mankind.

\section{Controlling light pollution}
A new challenge to protection of observatories from light pollution is coming from a widespread switch towards blue-rich artificial light sources, whose light energy emission peaks near 450 nm, away from sodium-based light sources, which peak near 590 nm.  At most observatory sites, Rayleigh scattering is the dominant mechanism that brings artificial light into the telescope, and this scattering is strongly wavelength dependent -- blue light scatters much more efficiently than amber (sodium) light. 
The emission from blue rich light sources peaks at wavelengths where the human eye is not very sensitive, but the natural night sky is very dark and astronomical detectors very sensitive indeed. 
\smallskip

Recent advances in metal halide, induction, fluorescent and particularly light emitting diodes (LEDs) have made these blue rich light sources competitive with sodium lights in terms of energy efficiency (lumens per Watt), and some LEDs are now more efficient than sodium lamps. Some municipalities are selecting the most energy efficient LEDs, which are very blue rich, despite the harm that they do to the night sky; blue rich white light is also harmful to many species of animals, and affects the circadian rhythm in humans and animals. 
\smallskip

Several approaches can help to minimize the impact of LEDs and other blue-rich white light sources on astronomy. These include 3 methods, of which nos. 2 and 3 result in much less impact on observatories than method 1:

\begin{itemize}
\item 1.	Use of low correlated color temperature light sources -- CCT $<$ 3,000K is strongly recommended.  More energy efficient ``warm white'' LEDs, which use a phosphor to convert some of their blue light to longer wavelengths, have recently become available, and metal halide, induction and fluorescent lamps and LEDs are all readily available with low CCT.
\item 2.	Filtering out the blue light.  On the island of Hawaii, where Mauna Kea Observatory is located, LED light sources have been installed that use a filter that absorbs nearly all light below 500 nm.  
\item 3.	Use of a carefully crafted mix of LEDs.  For example, a mix of 3 amber LEDs and one low CCT white LED produces a light with adequate color rendition, low CCT and minimal blue light.
\end{itemize}
\smallskip

On the positive side, light from LED light sources can be much more efficiently directed, and most lighting tasks can efficiently be performed using fully shielded light sources.  So a lighting task can be achieved using many less total lumens using LEDs.  Use of fully shielded light sources has been shown to be of paramount importance for the protection of observatories.
We have been cooperating with the {\it International Dark Sky Association} (IDA) in some aspects of this.
\smallskip

Commission 50 members have increased their presence at meetings of the 
{\it Commission internationale de l'\'eclairage} (CIE), the professional 
organization of lighting engineers, and participated in a number of technical 
committees. The IAU liaison with the CIE is 
Elizabeth Alvarez del Castillo, and David Galadí-Enríquez is our representative
on the new CIE Lighting Quality and Energy Efficiency Network, of which the IAU 
is one of the participating organizations.
At the CIE's 2009 Interim Conference, Connie Walker explained 
how light emitted near the horizontal contributes significantly to sky 
glow (Luginbuhl 2009), while Richard Wainscoat showed why the 
newer blue-rich white light sources are especially harmful to astronomical 
research (see the Section above). In addition to technical committee meetings, CIE 
Divisions 4 and 5 held a special workshop on LEDs, at which Francisco 
Javier Díaz Castro and Elizabeth Alvarez represented the IAU. In spring 
2010, Ferdinando Patat gave an invited talk at a special CIE conference on 
energy efficient lighting. In fall 2010, we had standard representation at 
the CIE Divisions 4 and 5 meetings. C50 members David Galadí-Enríquez and 
Ramotholo Sefako attended the CIE Quadrennial in 2011. Emphasis has been put
this past triennium on raising awareness of the 2 main issues on which we gave 
talks at the 2009 CIE Conference.
\smallskip

Commission 50 has strengthened its links with the {\it International Dark Sky Association} (IDA). 
Connie Walker chairs the IDA Education Committee and has been nominated as member of the IDA Board of Directors; she was awarded the 2011 IDA Hoag-Robinson Award.
\smallskip

One way the legacy of the 2009 IAU/UNESCO International Year of Astronomy's Global Cornerstone Project ``Dark Skies Awareness'' has continued is through the Global Astronomy Month's ``Dark Skies Awareness'' programs (Connie Walker, chair). Global Astronomy Month (GAM) has taken place in April over the last two years and is continuing this coming year. One of the main ``take-away'' messages from GAM is why we should preserve our dark night skies. Several dark skies events and activities are being held worldwide on behalf of GAM to promote public awareness on how to save energy and save our night sky. Specific programs include:
\begin{itemize}
\item GLOBE at Night: an international citizen-science campaign to raise public awareness of the impact of light pollution by inviting people to measure their night sky brightness and submit their observations to a website. The campaign has run for two weeks each winter/spring for the last six years. People in 115 countries have contributed 66,000 measurements, making GLOBE at Night one of the most successful light pollution awareness campaigns (Connie Walker, director). 
\item Dark Skies Rangers: an environmental/astronomy-based program that includes student participation in GLOBE at Night. Students learn the importance of dark skies and immerse themselves in, e.g. activities that illustrate proper lighting technology and what light pollution's effects are on wildlife (Connie Walker, director). 
\item International Dark Sky Week: with activities, events and info on how to light more responsibly (Connie Walker, director), done with the IDA.
\item International Earth \& Sky Photo Contest: inviting landscape astrophotographers to capture the beauty of the night sky and/or light pollution's effect on it, done with The World At Night and NOAO.
\item World Night in Defense of Starlight (April 20): recognizes the importance of a pristinely dark sky as it relates to our culture, done with the Starlight Initiative.
\item One Star at a Time: creates accessible public places to observe a dark night sky. 
\item Dark Skies Awareness: with 10 minute audio podcasts on the ``Dark Skies Crusader'' and the effects of light pollution on wildlife, energy, health and astronomy, done with NOAO.
\item Astropoetry: on dark skies awareness.
\end{itemize}
\smallskip

Commission 50 and other IAU members participated in the 9th, 10th and 11th 
European Symposium for the Protection of the Night Sky (Dark Sky Symposium), which were held in 2009, 2010 and 2011, respectively.
The aim of the Symposium is to address issues of light pollution, its causes, its negative effects and possible remedies. Some of the specific issues have been
measuring methods of light pollution, the quality of dark sky parks, efficient dark sky friendly lighting and environmental impact (e.g. on insects, humans) of modern light sources (e.g. LEDs).
We presented the activities of our Commission, biological studies and numerous Dark Skies
Awareness Programs.
\smallskip

Commission 50 endorsed a proposal submitted by Richard Green for a special session on light pollution at the upcoming 2012 IAU General Assembly in Beijing, which was later merged with another proposal submitted by Beatriz Garcia of Commission 46 (Astronomy Education and Development), to be ultimately approved as SpS17 -- Light Pollution: Protecting Astronomical Sites and Increasing Global Awareness through Education. 
The progam will cover a combination of education and outreach topics, and more technical ones focused on astronomical site protection and the potential of spectral encroachment of new types of light sources in the blue.

\section{Controlling Radio Frequency Interference}
The main goal is to represent the interests of the international astronomical community in matters concerning the protection of radio frequencies allocated to the Radio Astronomy Service and minimize interference to our scientific observations and measurements.
\smallskip

The {\it International Telecommunication Union} (ITU) is an international organization under the United Nations, which establishes and maintains international rules on frequency use. 
IAU interaction with the ITU is primarily made via the {\it Scientific Committee on Frequency Allocations for Radio Astronomy and Space Science} (IUCAF). This committee is composed of representatives from the IAU, URSI ({\it International Union of Radio Science}) and COSPAR ({\it Committee on Space Research}), and it operates as an inter-disciplinary committee of the {\it International Council for Science} (ICSU).
\smallskip

Three Commission 50 officers are members of IUCAF: Masatoshi Ohishi, Tasso Tzioumis and Wim van Driel. Masatoshi Ohishi, the IUCAF chair, served as liaison between the IAU and the ITU, and Tasso Tzioumis chaired the Working Group on Radio Frequency Interference of IAU Division X until August 2009.
\smallskip

The interests and activities of IUCAF range from preserving what has been achieved through regulatory measures or mitigation techniques, to looking far into the future of high frequency use, giant radio telescope use and large-scale distributed radio telescopes. Current priorities, which will certainly keep us busy through the next years, include the use of powerful radars and satellite down-links close in frequency to the radio astronomy bands, the coordination of the operation in shared bands of radio observatories and powerful transmissions from downward-looking satellite radars, the possible detrimental effects of ultra-wide band (UWB) transmissions at around 24/79 GHz regions and high-frequency power line communications (HF-PLC) on all passive services, the scientific use of the 275 to 3000 GHz frequency range, and studies on the operational conditions that will allow the successful operation of future giant radio telescopes.
\smallskip

In the period 2009-2011, IAU representatives on IUCAF participated in a total of 14 international meetings, mainly of various ITU Working Parties and Study Groups. At the ITU, the work of interest to the IAU was focused on the relevant agenda items that were adopted in 2007 for the
World Radiocommunication Conference in 2012, WRC-12, as well as on the creation and maintenance of various ITU-R Recommendations and ITU-R Reports. They were also active in the organization of
the 3rd IUCAF Summer School on Spectrum Management for Radio Astronomy, which was held in 2010, at the National Astronomical Observatory of Japan (NAOJ) in Tokyo.
\smallskip

The WRC-12 agenda item which is most relevant to radio astronomy concerns the use of the radio spectrum between 275 and 3000 GHz ($\lambda$$\lambda$ 0.1-1 mm), a frequency range used for observations of important spectral lines and continuum bands. Significant infrastructure investments are being made for the use of these bands, such as the Atacama Large Millimeter/submillimeter Array (ALMA), a facility currently under construction in northern Chile.
\smallskip

No frequency allocations for the use of this frequency range will be made at WRC-12, but the radio astronomy community was tasked to identify a list of specific bands of interest. This list was established in close collaboration with the IAU Working Group on Important Spectral Lines (chaired by Masatoshi Ohishi), and in 2010 a new ITU-R Recommendation RA.1860 was adopted on preferred frequency bands for radio astronomical measurements in the range 1-3 THz. Another recent Report, ITU-R RA.2189, showed that the frequency range above 1000 GHz can be used by both passive (i.e., receive-only -- such as radio astronomy) and active (transmitting) radio services with little possibility of interference.
\smallskip

Power Line Communications (PLC) utilizing the 2-30 MHz frequency range is a technology to send electrical signals for communication purposes through power lines that were designed and installed to carry current at 50/60Hz only. There has been serious concern that the electromagnetic field radiated by power lines may cause harmful interference to radio astronomical observations. Radio astronomers submitted several documents containing measurement results of actual harmful interference from PLC and theoretical analyses to the ITU-R, which were included in ITU-R Report SM.2158. 

\section{Closing remarks}
Our Commission continues to recognize that more professional astronomers, especially IAU members, need to become involved in efforts to reduce light pollution and radio interference -- see also the 2009 IAU Resolution B5 referred to in the Introduction.

\vspace{3mm}
 
{\hfill Wim van Driel}

{\hfill {\it president of the Commission}

\end{document}